# *Hybridization of Magnetism and Piezoelectricity for an Energy Scavenger based on Temporal Variation of Temperature*

Louis Carlioz[1,2] (louis.carlioz@imag.fr), Jérôme Delamare[2],
Skandar Basrour[1] and Guylaine Poulin[2]

[1]Micro and Nano Systems Group - TIMA, 46 avenue Félix Viallet, 38000 Grenoble, FRANCE
[2]MagMems– G2ELab, 961 rue de la Physique, BP46, 38402 St Martin d'Hères Cedex, FRANCE

*Abstract -* **Autonomous microsystems are confronted today to a major challenge: the one of energy supply. Energy scavenging, i.e. collecting energy from the ambient environment has been developed to answer this problematic. Various sources have already been successfully used (solar, vibration). This article presents temporal variations of temperature as a new source of exploitable energy. A brief review will take place at the beginning, exposing the different approaches used in the past. Then we will focus our attention on hybridization of magnetism and piezoelectricity. A new kind of thermal generator is proposed and a preliminary model is exposed. Conclusions will be drawn on the suitability of this prototype and the improvements that are needed to increase its potential.**

## I. INTRODUCTION AND MOTIVATION

With the emergence of MEMS sensors, of RF communications and of ultra low power electronics, integrated systems consuming in the range of tens of µW are now a reality. Such "intelligent" sensors are planned to be used in networks, composed of numerous tiny and cheap nodes. Those monitoring arrays will probably take an important pre-eminence in the near future for surveying buildings (temperature, humidity, seismic activity…), checking car parameters (temperature, pressure…) or health status when embedded in clothes.

One limitation of the current approach is that such systems are powered by a battery, bulky and of limited lifetime. This leads to a real nightmare when there is need for batteries replacement or disposal. Thus for ecological and economical reasons, a promising alternative is to render those nodes autonomous, for example by scavenging free energy from the ambient environment. This field has encountered an ever growing interest during the last decades and various harvesters have been prototyped. Such prototypes exploiting energy deriving from solar light, mechanical vibrations or radio emissions have been described and researched thoroughly before [1].

We took an interest in energy scavenging based on a change of temperature along time (and not along the distance, as in thermoelectricity for instance). Two main reasons motivated this choice: the first one being that temperature temporal variation has not been as deeply researched as other scavenging methods so far. The second one is the need for a method working when the others fail (no light, no vibrations). Since literature is nearly nonexistent about this specific topic, our first task was to select the most promising properties for a harvester.

## II. BRIEF REVIEW OF THERMAL METHODS FOR ENERGY SCAVENGING

When speaking of harvesting energy from a change of temperature, several physical properties can be thought of and exploited. The most direct way is pyroelectricity which converts temperature variation in electricity (Figure 1). Pyroelectricity can be seen as a pendant of piezoelectricity since quite often both aspects are present; the only difference is that instead of applying a pressure to generate charges, you change the temperature of your sample. Pyroelectricity is mostly dependent on the variation rate. Some measurements were done and for an increase of 10 °C in 20 s, the energy density was found to be 1 µW.cm$^{-3}$.

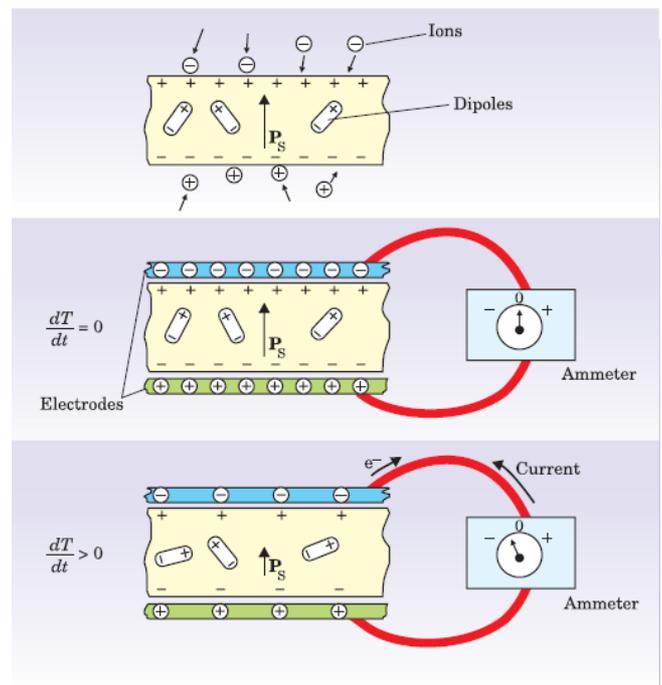

*Figure 1: Simple model of pyroelectricity from "Pyroelectricity: From ancient curiosity to modern imaging tool" by S. Lang*

Another approach is to use magnetic materials, since magnetism is always varying with temperature. One can think of using the change of magnetic state happening at the Curie temperature to create a difference of magnetic flux to be collected in a coil. The shifting of axis of magnetization (Figure 2) can also be a way to pursue.

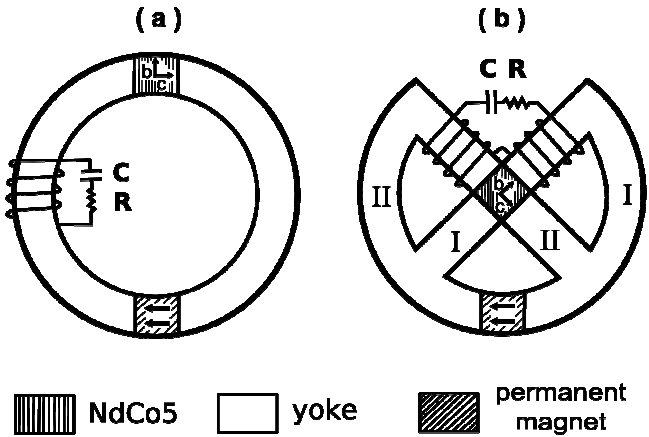

NdCo5    yoke    permanent magnet

*Figure 2: Direct use of magnetization reversal for electricity production [2]*

Such ideas have been implemented in the past and are working but gave an amount of energy too small to be significant [2]. Some investigations have been made with new magnetic materials like the ones used for magnetic refrigeration which show high "jumps" of magnetization, or with samples presenting "strange" rotation behavior. The conclusion was that those materials were either too expensive or too difficult to produce for use in cheap and easily made microsystems. Furthermore, they also exhibit a temperature rate dependency like pyroelectricity which makes them useless for small temperature variations.

This "time dependency" is probably the most important drawback to overcome in order to succeed in the creation of a thermal harvester. That's the reason why we try to develop a prototype which is as much independent as possible from time, thus for which a small temperature variation leads to a maximal electricity variation. Finally we made the choice to explore the hybridization of piezoelectricity and magnetism, both techniques which were so far used with mechanical excitation.

### III. PRESENTATION OF THE PROTOTYPE

In order to better apprehend how to limit the time dependence, let's start with a description of the macro scale harvester (Figure 3). It is composed of four different parts: the first one is a piezoelectric beam (bimorph) composed of a conducting metallic sheet sandwiched by two layers of piezoelectric material (PZT). There are consequently three electrodes and the possibility to connect the two piezoelectric layers either in parallel or in series. This beam is fixed on one end, and a classical NdFeB magnet is attached on the second end, acting both as a proof mass and as an actuator. It will act as an actuator since it will be attracted to the two remaining parts of the system. Those parts are identical; both are sheets of FeNi compounds (same composition) which have a specific thermal behavior. Our sample is a soft ferromagnet with a Curie temperature fixed at 45 °C (but this value can be tuned). The consequence is a relative permeability equal to 1 above $T_C$ and increasing when the temperature decreases.

If we were to exploit this phenomenon directly (via a coil wound around the NiFe), the same problem, as previously, would arise. Instead, we use the magnetic attraction and the mechanical stiffness of the piezoelectric beam to counterbalance each other. Since the magnetic force is non linear (varying in $1/r^4$ where r is the distance between the magnet and one FeNi sheet [3]), it is possible to create some kind of thermal threshold.

Below the threshold (Figure 4), the beam will stick to one of the two sides of FeNi, the magnetic force $F_M$ being preeminent with regards to the mechanical spring force $F_P$ of the beam. Conversely, as soon as $F_M < F_P$, the beam is getting attracted toward the equilibrium position in the middle of the two FeNi sheets. Since the magnetic force is non linear, when the gap increases, the magnetic force decreases drastically. The end of the beam is in first approximation only submitted to the mechanical force which leads to attenuated oscillations.

The magnetic modification of the FeNi is reversible, which means that when the temperature decreases below a second threshold, the beam will be attracted back towards the FeNi. At first the two FeNi compounds will tend to pin the beam due to compensating forces. Then at some point, since it's a disequilibrium state, the beam will move towards the top or bottom position. Once the movement is initiated, the closer the beam will be to that position, the stronger it will be attracted. Consequently, the magnet will nearly "jump".

The tuning of the system has to be done through a careful design of its geometry. This cannot be done manually by building several different harvesters, thus a modeling of the system has to be computed. The best way to harvest energy has to be investigated as well: indeed it is possible to use the deformation of the piezoelectric beam and / or some coil collecting the variation of magnetic flux coming from the fast moving magnet. Both approaches modify the forces acting on the beam, and calculations are needed in order to determine the best solution.

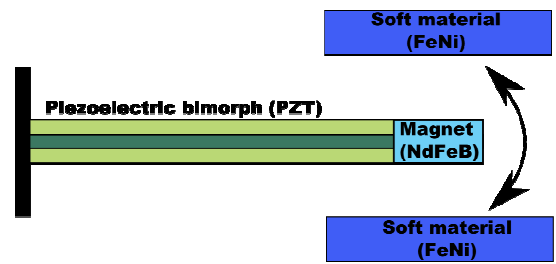

*Figure 3: Operating principle of the hybrid prototype. The magnet end is either attracted towards the FeNi or oscillates down to equilibrium position shown here*

### IV. MODEL OF THE HARVESTER

A simplified version of the system has been developed under the Matlab / Simulink environment, in order to verify the validity of this approach. At first, the magnetic force was approximated to its $1/r^4$ term and the beam was limited to its mechanical part (spring constant plus damping coefficient). We assert that for some geometry and coefficients, it was indeed possible to achieve the expected behavior (Figure 4).

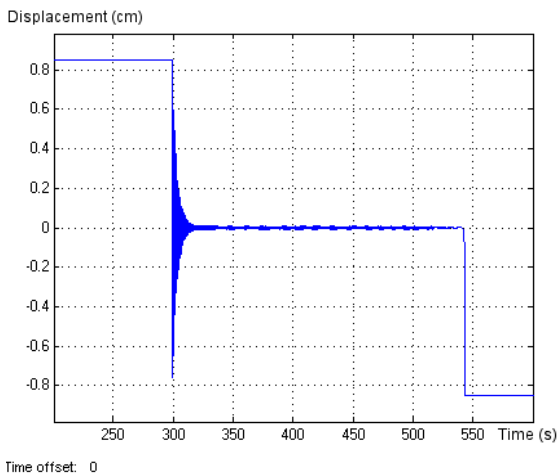

*Figure 4: Computed displacement of the magnet as a function of time (and thus of temperature, since the temperature is cycled around $T_C$ at 1°C/10s). The displacement is voluntarily limited to [-0.85, 0.85 cm]*

The model was further refined with the addition of the piezoelectric force deriving from the constitutive equations of piezoelectricity. For the moment, the complex losses have not been used.

Another addition under development is a more precise description of the magnetic part of the system. A traditional approach would be to use some finite element tool like Flux3D or Ansys, but since the magnetic state change at nearly every step (when above the threshold) this requires long calculation time.

We decide to use a numerical only approach through a special piece of software running on Matlab. This software, named Locapi, is able to compute first the effect of the magnet on the FeNi sheets (magnetization process) using the moment method (Figure 5) [4]. Then in a second step, it can derive the force applied by the FeNi on the magnet.

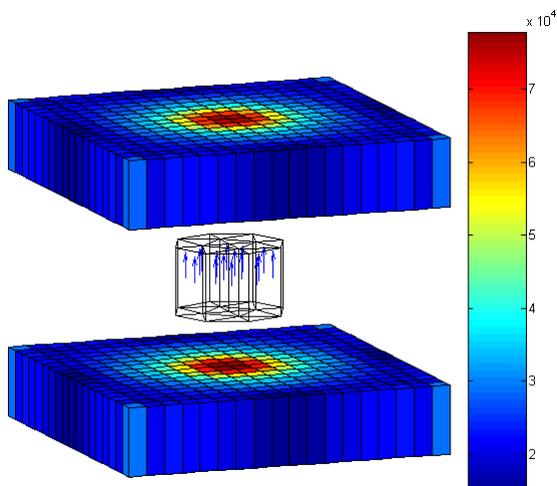

*Figure 5: Magnetization of the two FeNi sheets by the magnet in-between (3D simulation using Locapi)*

The main advantages of this software are its velocity (few minutes of computation per step) and the possibility to exploit directly the result under the Simulink environment, making it possible to have a direct analysis of the system.

## IV.    CONCLUSION AND FUTURE WORK

The model developed in this study will be used to design a real micro power generator based on thermal variation. It is still quite simple but the behavior computed was coherent with the one we expected.

Further refinements will include some routines to integrate directly the results from Locapi and thus have a self-contained model. Some efforts have to be done also to take into account the complex losses of the piezoelectric effect. Some optimization will also be carried out in order to find the best geometry leading to the maximum energy harvestable.

This work will be lead to the realization of a real generator able to harvest energy from temperature variation. Once build, we will focus our attention to scaling effects and will conclude on the efficiency of a micro scale generator.

### ACKNOLEDGEMENT

The authors would like to thank the INPG (Grenoble Institute of Technology) for funding (BQR microsources d'énergie).

### BIOGRAPHY

Louis Carlioz was born in 1984 in Bonneville, France. He is an electronics engineer and received a joint European master degree in micro and nano technologies in 2006 from the INPG (France), Polito (Italy) and EPFL (Switzerland). He is now a PhD student working on energy scavenging at G2Elab and TIMA in Grenoble.

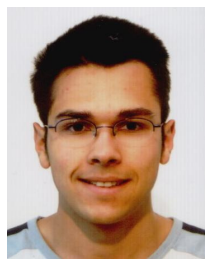